\documentclass{article}

\PassOptionsToPackage{numbers, compress}{natbib}

\usepackage[preprint]{neurips_ACA_2025}

\workshoptitle{\\ Workshop on Algorithmic Collective Action (ACA@NeurIPS 2025)}

\usepackage[utf8]{inputenc} 
\usepackage[T1]{fontenc}    
\usepackage{hyperref}       
\usepackage{url}            
\usepackage{booktabs}       
\usepackage{amsfonts}       
\usepackage{nicefrac}       
\usepackage{microtype}      
\usepackage{xcolor}         

\title{AI Workers, Geopolitics, and Algorithmic Collective Action}

\author{%
  Sydney Reis\\
  Responsible Technology Institute\\ 
  Department of Computer Science\\
  University of Oxford\\
  Oxford, UK\\
  \texttt{sydney.reis@wolfson.ox.ac.uk}\\
}

\begin{document}

\maketitle

\begin{abstract}

According to the theory of International Political Economy (IPE), states are often incentivized to rely on rather than constrain powerful corporations. For this reason, IPE provides a useful lens to explain why efforts to govern Artificial Intelligence (AI) at the international and national levels have thus far been developed, applied, and enforced unevenly. Building on recent work that explores how AI companies engage in geopolitics, this position paper argues that some AI workers can be considered actors of geopolitics. It makes the timely case that governance alone cannot ensure responsible, ethical, or robust AI development and use, and greater attention should be paid to bottom-up interventions at the site of AI development. AI workers themselves should be situated as individual agents of change, especially when considering their potential to foster Algorithmic Collective Action (ACA). Drawing on methods of Participatory Design (PD), this paper proposes engaging AI workers as sources of knowledge, relative power, and intentionality to encourage more responsible and just AI development and create the conditions that can facilitate ACA. 
   
\end{abstract}

\section{Introduction}

According to International Relations (IR) scholar Kenneth Waltz, “an agent is powerful to the extent that he affects others more than they affect him” \cite{waltz_theory_1979}. In recent years, companies at the forefront of Artificial Intelligence (AI) development are gaining more geopolitical\footnote{The "critical" approach and definition to geopolitics will be used in this position paper, without the "critical" foregrounding, as is typically done amongst scholars of critical geopolitics. Critical geopolitics is defined by Klaus Dodds as "focus[ing] more on the role of discourse and ideology... rather than conceptualize geography as deterministic, the geographical is seen to be rather more fluid and subject to interpretation. If classical geopolitics focuses on territory, resources, and location, critical approaches focus on how the interactions between the human and physical produces ‘geopolitics.’"\cite{dodds_geopolitics_2007}} power and influence\cite{schwartz_exclusive_2025, lehdonvirta_compute_2024, gill_whoever_2020, the_white_house_executie_office_winning_2025, pavel_ai_2023, chesterman_silicon_2025, srivastava_ai_2024, bomont_challenging_2025, quansah_deepseek_2025, srivastava_algorithmic_2023} with some scholars indicating that this power and influence represents a new form of empire.\cite{hao_empire_2025, _empire_2025} These trends have several implications. The International AI Safety Report, a landmark document in AI safety published after the 2023 AI Action Summit\cite{uk_government_chairs_2023}, acknowledges the "strong competitive pressure[s] which may lead [governments and AI companies] to deprioritize risk management."\cite{bengio_international_nodate} At the same time, several academics have recorded how Western technology companies and authoritarian regimes have overlapping interests which result in material or inadvertent support for authoritarian practices such as surveillance and censorship.\cite{jack_feminist_2021,conduit_digital_2025,feldstein_global_nodate}

Meanwhile, Science and Technology Studies (STS) has long held that the “stamp of conscious or unconscious human choice” is marked at every point of a given technology’s design, development, and dissemination process.\cite{bijker_social_2012, jasanoff_states_2006} Therefore, while AI is emerging as a geopolitical artefact,\cite{wasi_generative_2025} the companies and workers that develop AI\footnote{Throughout this paper, the term \textit{AI workers} is used. Though this paper acknowledges the various "invisible" aspects of AI labour,\cite{stilgoe_ghost_2022, altenried_platform_2019}, which is often unseen, undervalued, and performed by workers in the gig-economy and global South (e.g. data annotation, content moderation, and platform maintenance) for the purposes of this paper the term \textit{AI worker} should be understood as AI researchers and engineers within corporate and academic labs, following the literature on activism in the AI community.\cite{belfield_activism_2020, tan_unlikely_2024}} become actors of geopolitics.\cite{wasi_generative_2025} This is both because of their political agency\cite{belfield_activism_2020, jack_feminist_2021} and because of their privileged positions as the architects of AI systems\cite{csernatoni_myth_2025} that take on their values.\cite{bijker_social_2012, jasanoff_states_2006, wasi_generative_2025} The geopolitical influence of the AI company and AI worker has been exacerbated by a retreat of formal technical governance in recent years, especially in the United States\cite{the_white_house_executie_office_winning_2025, vance_remarks_2025, depirri_tale_2025, wei_how_2024, the_white_house_initial_2025} where the most powerful AI models are developed.\cite{kulothungan_towards_2025}  

Recent literature that examines the intersection between AI companies and geopolitics sometimes presents enhanced governance as a potential solution.\cite{pavel_ai_2023, wasi_generative_2025} However, these governance proposals often overlook how dynamics within the international political economy can weaken the incentives of governments to constrain large and powerful corporations.\cite{strange_retreat_2000, farrell_weaponized_2019} When considering the impact of IPE, it becomes necessary to complement AI governance mechanisms with other forms of resistance. Csernatoni et al. encourage reflection on AI-enabled power redistributions as a form of resistance.\cite{csernatoni_myth_2025} While this type of reflection on power relations is a useful academic exercise in itself, the resulting ideas and conversations likely has limited reach and influence on the AI development that occurs in industry. Therefore, reflection interventions should be developed for, and ideally in conversation with, the workers that drive AI innovation. In this new paradigm of AI company power concentration and geopolitical influence,\cite{schwartz_exclusive_2025, lehdonvirta_compute_2024, gill_whoever_2020, the_white_house_executie_office_winning_2025, pavel_ai_2023, chesterman_silicon_2025, srivastava_ai_2024, bomont_challenging_2025, quansah_deepseek_2025, srivastava_algorithmic_2023,hao_empire_2025,_empire_2025} bottom-up interventions at the tech worker level can and should complement AI governance options. Critical reflection\footnote{Critical reflection involves taking into consideration aspects beyond the immediate context, for example moral and ethical issues and wider socio-historical and politico-cultural contexts.}\cite{ward_reflection_2004, smith_reflection_1995, fleck_rating_2012, freire_pedagogy_2017} amongst AI tech workers can contribute to building critical mass, encouraging collective action, and eventually, resistance. Algorithmic Collective Action (ACA) literature should examine and engage AI workers as potential agents of change within powerful corporations. Participatory Design (PD) literature offers time-tested guidance for creating tools and technologies that serve the populations that they are designed for,\cite{bodker_participatory_2018,qi_participatory_2025,gregory_scandinavian_nodate} while Human-Computer Interaction (HCI) literature is replete with examples and tools that can be adapted to engage AI workers to reflect on,\cite{baumer_reflective_2015, bentvelzen_revisiting_2022} and, if deemed necessary, resist, AI company power concentration and potential negative geopolitical impacts of the technologies they contribute to the development or proliferation of. 

This position paper builds on existing work\cite{schwartz_exclusive_2025, lehdonvirta_compute_2024, gill_whoever_2020, the_white_house_executie_office_winning_2025, pavel_ai_2023, chesterman_silicon_2025, srivastava_ai_2024, bomont_challenging_2025, quansah_deepseek_2025, srivastava_algorithmic_2023,hao_empire_2025,_empire_2025} to argue that AI workers are emerging geopolitical actors amidst the increasing geopolitical influence of AI companies. The geopolitical implications of AI requires “governments [to] look beyond traditional regulatory techniques to influence AI developments.”\cite{pavel_ai_2023} Considering this, there should be bottom-up tools that can facilitate ACA built by and for contemporary AI workers. In a recent publication, Csernatoni et al. conclude that the influence of AI on reshaping geopolitics necessitates “critical reflection.”\cite{csernatoni_myth_2025} HCI literature provides examples of technologies for facilitating critical\cite{lee_conversational_2025, lee_we_2023, elsayed-ali_responsible_2023} and worker\cite{ortiz_enabling_2018, kocielnik_designing_2018} reflection, while the design methodology of PD offers a blueprint to ensure that any such tools are AI worker informed.

This paper first outlines ACA, IPE, the connections between the two disciplines, and opportunities for further research. It will then advance the argument that AI workers are geopolitical actors capable of geopolitical change. Finally, it proposes a PD-informed approach for engaging AI workers in critical reflection and collective action. This work advances the academic discourse in ACA in three main ways. First, by using IPE literature to argue that bottom-up initiatives at the AI worker level can complement the goals of AI governance and ACA. Secondly, by positioning AI workers as agents of change with potentially far reaching impacts for ACA given the high-profile nature of their work, large user-base, and positionality. Lastly, it proposes PD-informed methods to empower AI workers to critically engage with geopolitics and ACA.

\section{Background and related work}

This section reviews relevant literature in ACA and IPE, while advocating for the broader inclusion of geopolitics in the ACA literature. 

\subsection{Algorithmic Collective Action}

Economics and Sociology have studied how groups of people with common interests come together to impact markets and politics through “collective action.”\cite{olson_introduction_1971, marwell_critical_1993} Notably, Marwell and Oliver introduced the theory of critical mass\cite{marwell_critical_1993}, arguing that a “minimum number of people or money” is required to spur collective action.\cite{marwell_critical_1993} Postmes and Brunsting, Turner et al., and Li et al., pioneered early work at the intersection of collective action and HCI, especially through the convening power of the internet.\cite{postmes_collective_2002, turner_picturing_2005, li_out_2018} In recent years, what has been termed Algorithmic Collective Action has since taken shape in Machine Learning (ML) literature.\cite{kulynych_pots_2020, hardt_algorithmic_2023, ben-dov_fairness_2025, baumann_algorithmic_2025, solanki_crowding_2025} ACA and its early iterations largely focused on how communication technologies, and later, ML, facilitates labour rights, fairness, and platform dynamics, especially for platform users.\cite{chesterman_silicon_2025} Sigg et al.'s living list of ACA efforts contains several references to collective action amongst platform workers who often face precarious employment and unfair labour practices, such as artists, gig, and contract workers.\cite{sigg_decline_2025}  

However, little work has been done on the potential for AI workers within corporate and academic labs and to engage in ACA, for instance, to provide the conditions necessary to facilitate ACA, or to design algorithms that might align business incentives with collective action. This is likely due to the perception that AI workers, because they are often highly-skilled, mobile, and afforded relative freedoms in their work, are not incentivized to engage in resistance, or address the economic, moral, social and political impacts of their technologies.\cite{olson_introduction_1971, kulynych_pots_2020} However, AI worker identities are multifaceted and there is a history of resistance in AI worker populations\cite{belfield_activism_2020}, even if this resistance seems to have lessened in recent years.\cite{gonzales_how_2024, abdalla_100000_2025} Given that AI companies are amalgamating power and exercising geopolitical will,\cite{schwartz_exclusive_2025, lehdonvirta_compute_2024, gill_whoever_2020, the_white_house_executie_office_winning_2025, pavel_ai_2023, chesterman_silicon_2025, srivastava_ai_2024, bomont_challenging_2025, quansah_deepseek_2025, hao_empire_2025} the ACA literature should more closely examine the role of AI workers \textit{in situ} to creating more just and equitable conditions for their technologies, platforms, and algorithms.

\subsection{The International Political Economy and the challenge of AI governance}

AI development and use causes negative "externalities" or harm at the individual, societal, cultural, and planetary levels.\cite{hernandez_mapping_2024} Increasing recognition of these harms has led to growing calls to govern AI. In recent years, AI governance has gained traction as an academic discipline\cite{batool_ai_2025, taeihagh_governance_2021}, policy imperative,\cite{ulnicane_framing_2021} political battleground,\cite{harris_regulating_2025, justo-hanani_politics_2022} and geopolitical tool\cite{roberts_global_2024, ishkhanyan_sovereignty-internationalism_2025}. Some academics argue that AI policy is often created in a vacuum and removed from democratic-style engagements with diverse sets of end-users.\cite{coeckelbergh_artificial_2025} Moreover, AI governance is widely considered to still be nascent and has several gaps in terms of enforcement and robust legal coverage.\cite{roberts_global_2024, the_united_nations_ai_advisory_body_govering_2024}. Other scholars have pointed out that AI governance lacks a consensus on values,\cite{correa_worldwide_2023} is applied unevenly,\cite{roberts_global_2024} and very often legs behind rapidly evolving technical realities.\cite{allen_roadmap_2025} These limitations reflect enduring challenges to the democratic governance of emerging science and innovation.\cite{stilgoe_developing_2013}

The IPE presents another limitation to AI governance, especially international AI governance. As a theory within the broader discipline of IR, IPE holds that Transnational Corporations wield power and uphold their interests in the international system in much the same that states do.\cite{strange_retreat_2000} In her seminal work, The Retreat of the State: The Diffusion of Power in the World Economy, Susan Strange states “where states were once the masters of markers, now it is the markets which, on many crucial issues, are the masters over the governments of states. And the declining authority of states is reflected in a growing diffusion of authority to other institutions and associations.”\cite{strange_retreat_2000} According to Strange, neoliberal policies of privatisation, the offshoring of manufacturing, and corporate tax avoidance have worked in tandem with technological know-how to accelerate these trends. Henry Farrell and Abraham Newman build on IPE to examine how states weaponize corporate infrastructure, supply chains and technologies to bolster national power.\cite{farrell_weaponized_2019} Farrell and Newman cite the surveillance power of the internet as an example of panoptic power, arguing that the US wields significant panoptic power against other states because of its near monopoly on internet information flows.\cite{farrell_weaponized_2019} Rather than cooperate with other states to regulate the internet, the US preferred to privatize and exploit internet technology to bolster international power.\cite{farrell_weaponized_2019}

AI companies have already begun making geopolitical decisions normally assumed to be within the remit of states.\cite{roulette_musk_2025, khanal_why_2025} Rather than trying to contain these companies through governance, IPE demonstrates that states have an incentive to remove any barriers that may jeopardize opportunities to bolster national power in the international system.\cite{strange_retreat_1996,farrell_weaponized_2019} This approach to technology governance is particularly apparent in the US and China, which by many measures, are the countries with the most sophisticated AI research and development.\cite{the_white_house_removing_2025, the_white_house_executie_office_winning_2025, roberts_chinese_2021, lundvall_chinas_2022, alberto_innovation_2020} For instance, several official government policies in both the US and China privilege AI innovation over governance.\cite{the_white_house_executie_office_winning_2025, webster_full_nodate} The realities of the IPE and strategic value of AI suggest that governance cannot be the only avenue to mitigate negative externalities. Bottom-up interventions inspired by ACA and HCI offer creative frameworks to de-risk AI and illuminate its geopolitical implications. 

\section{AI workers as geopolitical actors}

This section argues that viewing AI workers as agents of geopolitics can facilitate ACA. It first establishes the connection between geopolitics and AI workers. It then demonstrates that AI workers can and have previously enacted change within their places of work. Consequently, this section contends that AI workers are critical sites of intervention for enabling bottom-up collective action toward more responsible and just AI outcomes.

\subsection{The collective potential of AI workers}

There is an emerging body of literature that documents the geopolitical influence of AI companies.\cite{schwartz_exclusive_2025, lehdonvirta_compute_2024, gill_whoever_2020, the_white_house_executie_office_winning_2025, pavel_ai_2023, chesterman_silicon_2025, srivastava_ai_2024, bomont_challenging_2025, quansah_deepseek_2025, srivastava_algorithmic_2023,hao_empire_2025, _empire_2025} This literature is supported by recent partnerships between AI companies and states to develop government platforms, military technologies, and/or sovereign AI capabilities.\cite{hale_google_2025, hale_openai_2025, shepardson_amazon_2025, innovation_canada_2025} As architects operating on and within these systems, AI workers emerge as actors of geopolitics. 

In contrast to the artists, gig, and contract workers that have traditionally been studied in ACA literature, AI workers that engage in research and engineering tasks within corporate environments may have more specialized skills, mobility, and labour freedom.\cite{stephany_beyond_2025} Less attention may have been paid to these AI workers because they have historically been seen as “incompatible with labour activism.”\cite{tan_unlikely_2024, dorschel_reconsidering_2022, brophy_system_2006} However, the labour, mobility, and influence of AI workers can shape how and what systems get built in the first place and whose values they are aligned with. Moreover, AI workers have a history of activism and resistance, including in geopolitics.\cite{tan_unlikely_2024} In 2022, Google and Amazon workers published an anonymous letter\cite{anonymous_google_and_amazon_workers_we_2021} pressuring their employers to cancel a contract with Israel over fears of platforms and services being used against Palestinian people.\cite{koren_behind_2022} AI workers have political sway and collective will, and their collective actions vis-a-vis algorithmic infrastructures and their cites of use represents a research opportunity for ACA. 

AI workers can act as a collective that shape AI development towards or away from geopolitical harms.\cite{belfield_activism_2020, stephany_beyond_2025} Haydn Belfield notes that a coherent culture amongst AI workers and bargaining power due to the high demand and limited supply of AI talent have facilitated activist efforts.\cite{belfield_activism_2020, ray_google_nodate} At the international level, AI worker activism has featured in the debate over the international ban on Lethal Autonomous Weapons Systems.\cite{belfield_activism_2020} Notably, in 2018, Google employees pressured the company to back out of engagement with the US Department of Defence’s Project Maven.\cite{wakabayashi_google_2018}  

In recent years, there is some evidence from gray literature that big AI companies are engaging more in the military industrial complex with less resistance from workers.\cite{sun_stanford_2025} Amidst the backdrop of the so-called AI race with China in AI\cite{the_white_house_removing_2025} AI companies are increasingly proliferating the narrative that their platforms are “democratic.” \cite{openai_openais_2025, anthropic_request_2025} What benchmarks and parameters defines a so-called “democratic” platform are not specified. Amidst this shift, Mohamed Abdalla calls for “updating worker strategies for influencing corporate behaviour in an industry with vast social impact.”\cite{abdalla_100000_2025} In Gideon Kunda’s seminal ethnography of Silicon Valley "big tech workers", he observes that they are, above all, individuals with their own values who attempt to navigate the strongly imposed cultural and values of their workplace.\cite{kunda_engineering_1995} In the context of their increasing geopolitical agency, AI workers should have more opportunities to reflect on and collectively act in accordance with their own values, which may be in line with the goals of ACA. 

\section{Towards engaging with AI workers on geopolitics and ACA}

As observed by Hernández et al. in a recent paper surveying AI risks, “One of the biggest challenges in assessing and anticipating algorithmic harm has to do with limited evidence and difficulties related to observing the indirect manifestations of harm, as well as foreseeing its effects over time. Grasping these complexities involves nuanced and context-dependent understanding.”\cite{hernandez_mapping_2024} Despite these difficulties, cultivating a collective consciousness of the structural and geopolitical impacts of AI remains essential.\cite{hernandez_mapping_2024, sengers_reflective_2005} 

This paper proposes engaging AI workers through PD to co-develop tools or practices that help contextualise their work within their values and beliefs. PD emphasizes including user groups in the design process, guided by an ethos of empowering "people, in various communities and practices, to take control and partake in the shaping and delivery of technological solutions, processes of use, and future developments that matter to them and their peers."\cite{bodker_participatory_2018} The roots of PD are inherently worker centric, having originally emerged as an approach to help workers and unions navigate the introduction of computers in the workplace.\cite{bodker_participatory_2018} Although PD has historically been used to amplify the voices and needs of under-represented users in technology design, its application here is intentionally inverted. In this case, AI workers, as presented in this paper as a relatively privileged group, are engaged as participants capable of shaping tools and practices that could lay the groundwork for ACA. This process also provides a means for workers to interrogate their own geopolitical positionality and explore how it might be mobilized towards collective action within the AI systems they contribute to. In this way, the design process itself also becomes an intervention, fostering spaces for dialogue, critique, and ethical imagination.

This approach prioritises participation over artefact. The goal is not to prescribe a particular solution, but to co-create potential practices and/or prototypes that could enable AI workers to interrogate questions such as:

\begin{itemize}
	\item How do my design choices participate in geopolitics and global power dynamics?
    \item How does my company’s geopolitical positioning serve different state interests?
	\item What forms of resistance are possible from within? 
    \item Is it possible to situate this resistance within algorithms or AI systems themselves? 
\end{itemize}

\section{Conclusion}

This paper has argued that the IPE constrains AI governance and that, as a result, governance reform alone is unlikely to prevent many of the negative externalities produced by modern AI systems. Building on IPE, STS, and other recent gray and academic literature, this paper has shown that AI companies concentrate and wield power in ways that mirror state authority, and that the engineers and researchers who build these systems occupy a (geo)politically significant position within this context. Given identified constraints on governance, this paper proposes increased attention towards bottom-up interventions at the level of AI workers, who, for instance,  can help create the social and epistemic conditions necessary for ACA.

The paper makes three contributions. First, it extends ACA scholarship by centring AI workers as potential agents of change and resistance rather than limiting ACA analyses to unseen labourers such as content moderators and data labourers. Second, it connects IPE and ACA to explain why worker-centred interventions are both necessary and valuable when formal regulation is vulnerable. Third, it proposes a PD–informed research pathway for co-creating tools and practices with AI workers that can facilitate collective mobilization on geopolitical topics. 

In the context of the growing international importance of AI companies and retreat of formal AI governance, AI workers become an ever more important unit of intervention for bottom-up approaches to both mitigate AI's negative externalities and co-imagine more just uses. Ultimately, this paper calls for examining ACA as a distributed and participatory project that can be extended to non-traditional actors (or ones that might, at first blush, even be considered outright hostile to the aims of ACA) with geopolitics presenting as a timely and promising use case to this end. 

\bibliographystyle{plainnat}  

\bibliography{Bibliography.bib}

\end{document}